\documentclass[apjl]{emulateapj}
\usepackage{aas_macros}
\usepackage{natbib}
\usepackage{amsmath}
\begin{document}

\title{Magnetic Mixing in Red Giant and Asymptotic Giant Branch Stars}

\author{J.~Nordhaus\altaffilmark{1,2}, M.~Busso\altaffilmark{3}, G.~J.~Wasserburg\altaffilmark{4}, E.~G. Blackman\altaffilmark{1} and S.~Palmerini\altaffilmark{3}}
\altaffiltext{1}{Department of Physics and Astronomy, University of Rochester, 
Rochester, NY 14627 USA; nordhaus@pas.rochester.edu}
\altaffiltext{2}{Department of Astrophysical Sciences, Princeton University,
Princeton, NJ 08544 USA}
\altaffiltext{3}{Department of Physics, University of Perugia, via Pascoli, Perugia 06123, Italy and I.N.F.N., Section of Perugia}
\altaffiltext{4}{Division of Geological and Planetary Sciences, California Institute of Technology, Pasadena, CA 91125 USA}

\begin{abstract}
The available information on isotopic abundances in the atmospheres of low-mass Red Giant Branch (RGB) and Asymptotic Giant Branch (AGB) stars requires that episodes of extensive mixing occur below the convective envelope, reaching down to layers close to the hydrogen burning shell (Cool Bottom Processing).  Recently \cite{Busso:2007jw} suggested that dynamo-produced buoyant magnetic flux tubes could provide the necessary physical mechanisms and also supply sufficient transport rates.  Here, we present an $\alpha-\Omega$ dynamo in the envelope of an RGB/AGB star in which shear and rotation drain via turbulent dissipation and Poynting flux.  In this context, if the dynamo is to sustain throughout either phase, convection must resupply shear.  Under this condition, volume-averaged, peak toroidal field strengths of $\left<B_\phi\right>\simeq3\times10^3$ G (RGB) and $\left<B_\phi\right>\simeq5\times10^3$ G (AGB) are possible at the base of the convection zone.   If the magnetic fields are concentrated in flux tubes, the corresponding field strengths are comparable to those required by Cool Bottom Processing.

\end{abstract}


\section{Introduction}

In low-mass Red Giant Branch (RGB) and Asymptotic Giant Branch (AGB) stars, matter must circulate in episodes of extended mixing from the convective zone to the radiative region where nuclear processing can occur (the so-called Cool Bottom Processing; CBP; \citealt{Wasserburg:1995cj}; \citealt{Charbonnel:1998hi}; \citealt{Nollett:2003il}; \citealt{Herwig:2005eb}).  While the mechanism driving mixing remains under investigation, CBP predicts the chemical and isotopic evolution for a given bulk transport rate by utilizing the path integral of the nuclear reactions over the mass and independent of the mixing mechanism (\citealt{Boothroyd:1994to}; \citealt{Wasserburg:1995cj}).  The necessary transport rates and processing temperatures are constrained by RGB/AGB isotopic measurements and presolar meteoritic dust grain abundance ratios (\citealt{Choi:1998zt}; \citealt{Amari:2001xw}; \citealt{Nollett:2003il}; \citealt{Nittler:2005hq}).  Weak circulation ($\dot{M}\sim10^{-6}-10^{-8}$ $M_\odot$/yr) in which material is transported from the convective envelope to just above the H-burning shell, processed and then returned has had success in matching: (i.) $^{12}$C/$^{13}$C, C and N abundances in RGB stars, (ii.) $^{12}$C/$^{13}$C and N/O in AGB stars and (iii.) $^{26}$Al/$^{27}$Al, $^{18}$O/$^{16}$O, $^{17}$O/$^{16}$O and $^{12}$C/$^{13}$C in circumstellar dust grains found in meteorites that must come from AGB sources (\citealt{Wasserburg:1995cj}; \citealt{Nollett:2003il}; \citealt{Herwig:2005eb}).

Several physical origins have been proposed for the assumed mixing including: shear and thermohaline instabilities, meridional circulation and internal gravity waves (\citealt{Zahn:1992kb}; \citealt{Denissenkov:1996ye}; \citealt{Denissenkov:2003qm}; \citealt{Chaname:2005hs}; \citealt{Eggleton:2006fb}; \citealt{Charbonnel:2007wu}).  Purely rotation-induced mixing appears insufficient as isotopic changes are minimal during the RGB and lead to a quenching of s-process nucleosynthesis during the AGB (\citealt{Siess:2004oz}; \citealt{Palacios:2006qa}).  

Recently, it was proposed that buoyant magnetic flux tubes can induce the necessary transport rates required for Cool Bottom Processing (\citealt{Busso:2007jw}). Constraints on the field strengths for the required mixing rates were calculated while the fields were assumed to originate from a dynamo operating in the RGB and AGB interiors.  In particular,the proposed mechanism 
would provide a means of rapid transport of material to the convective mantle and may provide a solution to the problem of Li rich stars.

Magnetically mediated outflows resulting from dynamo amplification have been proposed as the origin of bipolarity in post-Asympotitc Giant Branch stars (post-AGB) and planetary nebula (PN) (\citealt{Pascoli:1997kc}; \citealt{Blackman:2001sy}; \citealt{Nordhaus:2006oq}; \citealt{Nordhaus:2007il}).  However, dynamo-induced shaping requires differential rotation.  This can be supplied by binary companions or sustained throughout the AGB phase by convective redistribution of angular momentum (\citealt{Nordhaus:2007il}).  Convection may resupply differential rotation (analogous to the $\lambda$-effect in the Sun) and sustain the magnetic fields throughout the RGB/AGB stages (\citealt{Rudiger:2004zv}; \citealt{Nordhaus:2007il}).  In particular, it may be that a weaker dynamo (supplying sufficient mixing) is sustained throughout the RGB/AGB phases, but that a binary companion is required to power bipolar jets in post-AGB stars \citep{Nordhaus:2008os}.

We investigate the origin of large-scale magnetic fields via a dynamical, $\alpha-\Omega$ dynamo operating at the base of the convection zone in an initially 1.5 $M_\odot$ main sequence star during the RGB and AGB phases.  Our stellar models are identical to those used in \cite{Busso:2007jw}, and allow for direct comparison.  In both phases, subadiabatic zones similar to the solar tachocline exist.  We allow a fraction of the turbulent energy cascade to resupply shear, thus sustaining the dynamo.  The field penetrates the shear zone and may transport material close to the H-burning shell.  We compare our results to the field strengths and penetration depths needed for adequate magnetic mixing in low-mass RGB and AGB stars.

\section{Magnetic Model}

\begin{figure}[h!]
\plotone{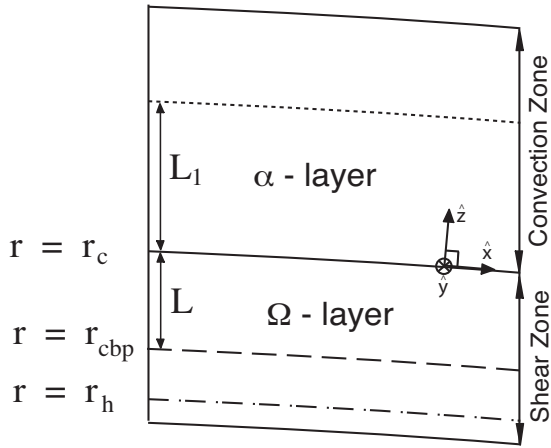}
\caption{Geometric representation of the stellar interior.  The poloidal field is amplified in the convective zone, while toroidal field is amplified in the shear layer.  The bottom of the convection zone is given by $r=r_c$ while the maximum depth required by Cool Bottom Processing is $r=r_{cbp}$.  The top of the H-burning shell is located at $r=r_h$.  \label{geometry}}
\end{figure}

A 2-D schematic of our 1-D model is presented in Fig. \ref{geometry}.  The convection zone extends from the stellar surface to the interface between the convective and radiative zones.  Convective twisting motions convert buoyant toroidal fields into poloidal fields through the $\alpha$-effect.  Below the convection layer, the differential rotation zone shears poloidal fields back into toroidal fields via the $\Omega$-effect.

We solve for the time evolution of dynamic quantities at the interface ($r=r_c$).  The rotation profile across the shear layer varies from $\Omega$ at the interface to $\Omega+\Delta\Omega$ at $r=r_c-L$.  Thus, $\Delta\Omega$ is a measure of shear in the differential rotation zone.  If $\Delta\Omega=0$, the system exhibits solid body rotation.  The average poloidal field $\left<B_p\right>$, and average toroidal field $\left<B_\phi\right>$, amplify from 1 G seed values until they are quenched through a drain of the available differential rotation energy.  We refer the reader to  \cite{Nordhaus:2007il} for a derivation of the precise mean-field equations solved.

\subsection{Shear Zone Penetration Depth}
To capture aspects of the 2-D, mean-field geometry within the framework of our 1-D time-dependent model, we employ two turbulent diffusion coefficients: $\beta_p$, corresponding to diffusion of the poloidal field (which grows primarily in the convective region) and $\beta_\phi$, corresponding to diffusion of the toroidal field (which is amplified in the differential rotation zone; see Fig. 1).  We also employ $\beta_\phi$ as the turbulent diffusion coefficient for the toroidal velocity.  

The convective region is highly turbulent and the differential rotation zone is weakly turbulent, therefore $\beta_\phi\ll\beta_p$.  The value of $\beta_\phi$ determines how far the toroidal component of the field can diffuse into the shear zone in a cycle period.  The further into the shear zone the toroidal field can penetrate, the greater the shear energy that can be extracted and utilized by the dynamo.   

The depth to which the toroidal field can diffuse into the shear layer in a cycle period, $\tau$, is defined as $\delta\simeq(\beta_\phi\tau)^{\frac{1}{2}}$.  The cycle period does increase in the dynamical regime, however, it does not change appreciably from its initial value.  We set $\delta=L$ so that the field diffuses to the depth required by Cool Bottom Processing ($r=r_{cbp}$).  This fixes $\beta_{\phi, RGB}\simeq5.6\times10^{14}$ cm$^2$ s$^{-1}$ and $\beta_{\phi, AGB}\simeq6.9\times10^{15}$ cm$^2$ s$^{-1}$ as the required transport rates are $\sim5$ flux tubes per year during the RGB and $\sim80$ flux tubes per year during the AGB \citep{Busso:2007jw}.  

\subsection{The Need For Convection Resupplying Shear}

As the magnetic field is amplified, differential rotation energy is drained to support its growth.  The back-reaction of field amplification on the differential rotation results in rapid termination of the dynamo ($\leq100$ yrs) from the initial available shear energy; see \cite{Nordhaus:2007il} for details.  In order to sustain the dynamo through an RGB/AGB lifetime, a constant differential rotation profile must be established.  This occurs in the sun as convection re-seeds shear through the $\lambda$-effect (\citealt{Rudiger:2004zv}).  Although it remains to be established if a similar effect occurs in evolved stars, by analogy to the solar case, we allow a fraction of the turbulent energy cascade to resupply shear.  Additionally, we keep the rotation at the interface fixed.  This is physically equivalent to storing the Poynting flux in the interface region.  If the field is trapped, the Poynting flux does not emerge from the layer and thus, does not spin down the envelope.  An RGB/AGB dynamo will remain stable when the following two conditions are met: (i.) convection resupplies shear (ii.) the Poynting flux is stored inside the envelope (\citealt{Nordhaus:2007il}).  

The differential rotation profile in the interior is unknown.  An estimate for the maximum sustainable shear that convection can resupply for a sufficiently rotating star is given by $\Delta\Omega_{max}=\left({M_c}/{M_{\Delta\Omega}}\right)^{1/2}\left({v}/{L}\right)$ where $M_c$ is the mass of the convective layer, $M_{\Delta\Omega}$ is the mass of the shear zone and $v$ is a typical convective velocity.  For the RGB phase, $\Delta\Omega_{max}\sim2.5\times10^{-5}$ s$^{-1}$ while $\Delta\Omega_{max}\sim5.5\times10^{-5}$ s$^{-1}$ for the AGB phase.  If the initial differential rotation were greater than $\Delta\Omega_{max}$, the magnetic field would extract the excess shear before relaxing to sustainable values.  In this case, an initial period of enhanced magnetic activity would be expected followed by steady dynamo action.  In \S 3, for suitable $\Delta\Omega$, we determine the fraction of turbulent cascade energy per unit time necessary to maintain a steady-state dynamo in our RGB/AGB model stars (e.g. \citealt{Nordhaus:2007il}).

\subsection{Flux Tubes vs. Average Fields}

Our calculations are for mean-fields and correspond to volume averages. When the ratio of average thermal to magnetic pressure
 $\beta={\frac{\left< P\right>}{ \left< B^2\right>/8\pi}}>1$ the magnetic field may be concentrated in flux tubes with an overall
 volume filling fraction $\zeta$, such that $\frac{\zeta}{1-\zeta}\simeq \frac{1}{ \beta}$ \citep{Blackman:1996rm}.  
When the magnetic pressure in flux tubes are in pressure balance with their exterior, the mean-field strength is then related to a flux tube field, $B_t$ by $\left<B^2\right> \sim \frac{\zeta}{1-\zeta}\left< B_t^2\right>$.  
Since the requirements from magnetic mixing \citep{Busso:2007jw} constrain the magnitude of flux tube fields to have $\zeta<1$, the mean field can be lower than the flux tube field, relaxing the demands on the strength of the dynamo and the required shear.

\begin{deluxetable}{llll}
\tablewidth{0.0in}
\tablecaption{Convective and Subconvective Parameters\label{table:par}}
\tablehead{
\colhead{Parameter}                  &
\colhead{Symbol}                    &
\colhead{RGB}           	&
\colhead{AGB}
}
\startdata
Base of conv. zone (cm)&  $r_c$  & $6.3\times10^{10}$  &   $5.4\times10^{10}$\\
CBP radius (cm)& $r_{cbp}$ & $3.4\times10^9$ &$1.9\times10^9$\\  
H-burning radius (cm) & $r_{h}$ & $2.0\times10^9$           &    $1.5\times10^9$     \\
Mass of conv. zone (g) &  $M_c$       &    $2.2\times10^{33}$              &   $1.1\times10^{33}$    \\
Mass of $\Omega$-layer (g)   &  $M_{\Delta\Omega}$   &     $1.0\times10^{31}$     &     $1.3\times10^{30}$    \\
Rot. at interface (s$^{-1}$) &	$\Omega(r_c)$	&	$3.0\times10^{-6}$	& $2.5\times10^{-6}$	\\
Diff. Rotation	(s$^{-1}$)	&	$\Delta\Omega$& $1.01\times10^{-5}$ & $1.67\times10^{-5}$	\\
\cutinhead{Model Results}
Toroidal Field (G)  &  $\left<B_\phi\right>$   &     2602$$      &     $4778$     \\
Poloidal Field (G) &  $\left<B_p\right>$   &     342$$      &     $596$    \\
Re-supply Rate & $f$ &	0.154 & 	0.075 \\
Vol. Filling Fraction &	$\zeta$ & 0.36	& 0.23\\
\cutinhead{CBP Requirements\footnote{From \cite{Busso:2007jw}}}
Flux Tube Field (G)	& 	$\left<B_{t,c}\right>$	&3450	&8600\\
Flux Tube Field (G) 	&	$\left<B_{t,cbp}\right>$	&$3.8\times10^5$& $5\times10^6$
\enddata
\end{deluxetable}

\section{Dynamo Amplified Magnetic Fields}

By requiring CBP transport rates, the necessary field strengths and buoyant velocities are calculated for a given stellar model.  We employ the same models as \cite{Busso:2007jw}, in which an initial 1.5 $M_\odot$ main sequence progenitor is evolved through the RGB and AGB.  For each phase, we determine what is required to sustain the dynamo and the corresponding saturated large-scale field strengths from our $\alpha-\Omega$ dynamo.  A summary of the model parameters can be found in Table \ref{table:par}.

\subsection{The RGB Case}

For our model RGB star,  the stellar mass is relatively unchanged from the main sequence ($M=1.49$ $M_\odot$).  The base of the convection zone is located at $r_c=6.3\times10^{10}$ cm while the maximum depth to which matter must circulate is $r_{cbp}=3.4\times10^9$ cm.  This implies a shear zone thickness of $L\simeq6\times10^{10}$ cm.  The mass of the shear layer and convective envelope for the RGB star are $M_{\Delta\Omega}=1.0\times10^{31}$ g and $M_c=2.2\times10^{33}$ respectively.  The rotation rate at $r=r_c$ is obtained by conserving angular momentum on spherical shells from a main sequence progenitor whose surface rotation velocity is $\sim$ 30 km s$^{-1}$ \citep{Kawaler:1987nx}.

Using CBP mass transfer rates, mixing depths and buoyant rise velocities, the necessary magnetic field strength at $r_c$ and $r_{cbp}$ can be constrained.  If the field is purely toroidal and in flux tubes that rise individually, then the average toroidal flux tube field required by CBP is $\left<B_{t,cbp}\right>\equiv\left<B_t\left(r_{cbp}\right)\right>=3.8\times10^5$ G.  At the base of the convection zone (assuming conservation of mass and magnetic flux), this corresponds to a field of $\left<B_{t,c}\right>\equiv\left<B_t\left(r_{c}\right)\right>\simeq3.5\times10^3$ G for a dipole geometry \citep{Busso:2007jw}.  

To compare, we calculate the mean field  at the base of the convection zone assuming that a fraction $f$, of the turbulent energy cascade rate resupplies shear \citep{Nordhaus:2007il}.  The value of $f$ is not known, nor is it established that convection resupplies shear in evolved stars.  Although our choice of $f$ is arbitrary, we purposely pick low values so that the requirements on the energy that convection must supply are reasonable. 

\begin{figure}[h!]
\plotone{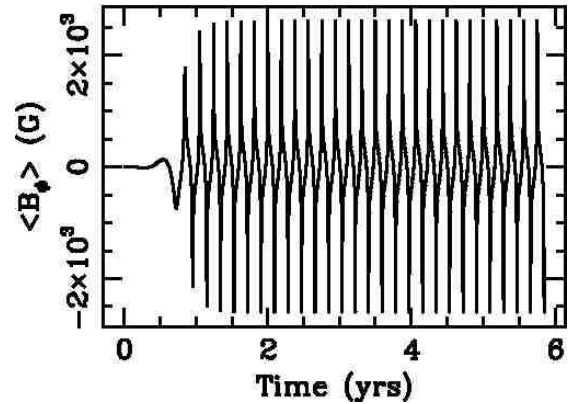}
\caption{Toroidal field as a function of time at the base of the convection zone for our model RGB star.  The field amplifies until it saturates. \label{rgb}}
\end{figure}

For the parameters in Table \ref{table:par}, the peak toroidal field at the base of the convection zone is $\left<B_\phi\right>=2.6\times10^3$ G (see Fig. \ref{rgb}).  In the RGB star, CBP requires approximately 5 buoyant flux tubes per year to reach the convective envelope.  This is comparable to the dynamo cycle period of  $\sim0.2$ yrs and suggests the necessary transport rates are plausible.  To sustain the dynamo throughout the RGB lifetime, $\sim15\%$ of the turbulent cascade energy must reinforce shear.  This is not a lower limit as smaller resupply rates will also sustain the dynamo, albeit with lower peak field strengths.

To compare our mean-fields with the flux tube fields necessary for CBP, we determine the volume filling fraction, $\zeta$, as detailed in \S 2.3.  For the RGB model, we find that $\zeta=0.36$.  However, it should be noted that the convective resupply rate is the dominant quantity as it sets the differential rotation and hence the peak field strengths.  A smaller $f$, implies a lower $\left<B_\phi\right>$ which requires a smaller $\zeta$ to match the required flux tube strengths.  Further work on how convection resupplies shear in evolved stars is warranted.

\subsection{The AGB Case}

The total mass of our AGB star has decreased from the initial 1.5 $M_\odot$ progenitor to $M=1.2$ $M_\odot$.  The base of the convection zone is located at $r_c=5.7\times10^{10}$ cm while the maximum depth required for CBP is $r_{cbp}=2\times10^9$ cm.  The shear zone thickness has slightly decreased from the RGB phase such that $L\simeq5.2\times10^{10}$ cm.  The mass of the convective layer and shear zone have decreased to $M_c=1.1\times10^{33}$ g and $M_{\Delta\Omega}=1.3\times10^{30}$ g respectively.  The rotation rate at the interface is given by $\Omega\left(r_c\right)=2.5\times10^{-6}$ s$^{-1}$ while the shear is slightly larger in the AGB phase.

\begin{figure}[h!]
\plotone{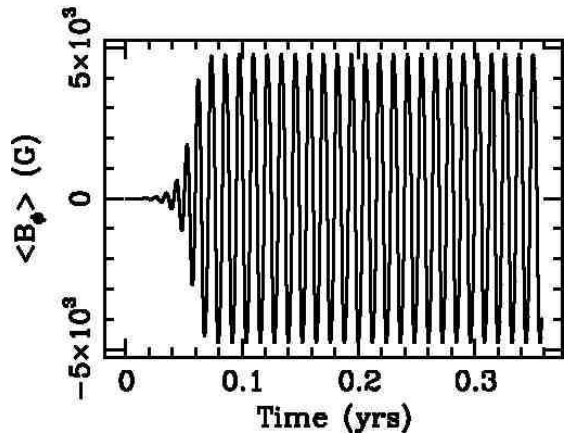}
\caption{Toroidal field at the base of the convection zone for our model AGB star. \label{agb}}
\end{figure}

A strong toroidal field is required at the CBP radius $\left<B_{t,cbp}\right>=5\times10^6$ G such that $\sim80$ flux tubes per year rise to the convective envelope.  At the base of the convection zone, the required flux tube field strength is $\left<B_{t,c}\right>\simeq8.6\times10^3$ G \citep{Busso:2007jw}.

From our dynamo calculations, the peak toroidal field at the base of the convection zone is $\left<B_\phi\right>=4.8\times10^3$ G.  Figure \ref{agb} shows the time evolution of the mean toroidal field in our model AGB star.    The cycle period is $\sim0.01$ yrs which is comparable to the required buoyant flux tube frequency.  

To sustain the dynamo, $\sim7\%$ of the turbulent energy cascade must resupply differential rotation.  In general, more convective energy is available in the AGB phase, and thus, the resupply rate necessary to support differential rotation is lower compared to the RGB star.  

Our mean fields require $\zeta=0.23$ to match CBP flux tubes at the base of the convective zone.  However, we note again that a lower $f$, will produce weaker peak field strengths and, in turn, a lower volume filling fraction.

\section{Conclusions}

\cite{Busso:2007jw} recently suggested that buoyant toroidal fields, may supply the necessary chemical transport rates to circulate material from near the hydrogen burning shell to the convective envelope required by Cool Bottom Processing in low-mass RGB and AGB stars. 

In this paper, we have shown how an $\alpha-\Omega$ dynamo operating in the interior of an RGB/AGB star could supply the fields necessary to accomplish the needed transport.  The fields diffuse in a magnetic cycle period to the depths required by Cool Bottom Processing.  If the dynamo is to sustain throughout either phase, convection must resupply differential rotation.  This can be accomplished if a fraction of the turbulent energy cascade reinforces shear ($\leq15\%$ for the RGB star; $\leq7\%$ for the AGB star).

For the model RGB star, mean toroidal field strengths of $\left<B_\phi\right>=2.6\times10^3$ G are sustainable at the base of the convection zone.  If the magnetic field is concentrated in flux tubes, the volume averaged fields, $\left<B^2\right>$, are related to the flux tube fields, $\left<B_t^2\right>$, by $\zeta<1$, such that $\left<B^2\right> \sim \frac{\zeta}{1-\zeta}\left< B_t^2\right>$.  For the model RGB star, $\zeta=0.36$ will satisfy CBP requirements.  For the model AGB star, stronger fields are possible with sustained toroidal field strengths of $\left<B_\phi\right>=4.8\times10^3$ G possible at the base of the convective zone.  This requires $\zeta=0.23$ to match CBP flux tube requirements.  

We have found that dynamo-produced fields provide a plausible mechanism for the physical origin of extra mixing in low-mass RGB and AGB stars.  However future work, in particular, realistic rotation profiles, the complex relationship between convection and shear, intermittent magnetic mixing, the dynamics of mass transport via flux tubes all need to be studied further.  Equally important is to independently constrain the value of the diffusion coefficients in the shear and convection zones to determine if they are consistent with what our model implies.  Finally, $\geq$ 2-D studies of this problem are needed.

\acknowledgements
J. N. and E. G. B. acknowledge support from SST grant GO-50134 and NSF grant AST-0406799.  M. B. acknowledges support from MURST (PRIN2006-022731).  G. J. W. acknowledges the support of DOE-FG03-88ER13851 and the generosity of the epsilon foundation and Caltech's contribution 9178(1123).

\end{document}